\documentclass{article}%
\usepackage{amsfonts}
\usepackage{aip}
\usepackage{amsmath}
\usepackage{amssymb}
\usepackage{graphicx}%
\providecommand{\U}[1]{\protect\rule{.1in}{.1in}}

\begin{document}

\title{Compressive Pressure, Spatial Confinement of Ions, and Adiabatic Heat
Generation in Binary Strong Electrolyte Solutions by an External Electric Field}
\author{Byung Chan Eu\\Department of Chemistry, McGill University\\801 Sherbrooke St. West, Montreal, QC\\H3A 2K6, Canada}
\date{The Date:
\today
}
\maketitle

\begin{abstract}
\setlength{\baselineskip}{20pt}In this paper, we make use of the exact
hydrodynamic solution for the Stokes equation for the velocity of a binary
ionic solution that we have recently obtained, and show that the
nonequilibrium pressure in an electrolyte solution subjected to an external
electric field can be not only compressive, but also divergent in the region
containing the coordinate origin at which the center ion of its ion atmosphere
is located. This divergent compressive pressure implies that it would be
theoretically possible to locally confine the ion and also to adiabatically
generate heat in the local by means of the external electric field. The field
dependence of pressure and thus heat emission is numerically shown and
tabulated together with the theoretical estimate of its upper bound, which is
exponential with respect to the field strength. It shows that, theoretically,
the Coulomb barrier between nuclei in the electrolyte solution (e.g., the ion
and a nucleus of the solvent molecule) can be overcome so as to make them fuse
together, if no other effects intervene to prevent it.

\end{abstract}

\section{Introduction}

\setlength{\baselineskip}{20pt}In a previous work\cite{euIonic} on the
solution of the Navier--Stokes (NS) equation---in fact, the Stokes
equation---for flow in electrolyte solutions subjected to an external electric
field, we have obtained exact solutions for the flow velocity of the medium
and the local nonequilibrium pressure as functions of the applied field
strength and other characteristic parameters of the system.\ The solutions
show that both the velocity of the countercurrent to the movement of ions
pulled by the applied external electric field and the (nonequilibrium)
pressure are divergent at the coordinate origin at which the center of the ion
atmosphere of an ion is positioned, whereas outside the region the pressure
profile is finite and can be positive. In particular, the excess
nonequilibrium pressure on the center ion not only is nonuniform in space, but
also becomes negative and divergent as the center ion is approached. This
means that the pressure on the center ion is compressive. Moreover, as will be
shown, the nonequilibrium part of the pressure is exponentially increasing in
magnitude with respect to the field strength as the field strength increases.

The divergent behavior of nonequilibrium pressure thus suggests, first of all,
that the compressional effect of the field gives rise to an emission of heat
(thermodynamic in origin), apart from the usual Ohmic heat due to resistance.
Thus, for example, it may explain the extraordinary heat generation which was
observed by Eckstrom and Schmeltzer during their conductance experiments
\cite{eckstrom} many years ago. Secondly, the spatially nonuniform and
divergent compressive pressure suggests a possibility of spatially confining
the ion with the external electric field together with the concerted help of
equally divergent countercurrent to the ions pulled by the field, while the
ions on the periphery of the ion atmosphere are delocalized and conduct
currents. This latter behavior was shown by the velocity
profiles\cite{eucond,archived} calculated from the solution of the Stokes
equation show similarly divergent profiles. This possibility of locally
confining ions by means of an external electric field seems quite interesting
and potentially rather significant. For example, at least theoretically, the
spatially confined ions can overcome the Coulomb barrier of nuclei if a
sufficiently high field is applied, and it would then be probable that they
might fuse together. In this paper, we would like to examine more closely the
compressive pressure with respect to the spatial position and also its
behavior with respect to the applied field strength.

We consider a binary strong electrolyte solution subjected to an external
electric field applied in the positive $x$ direction in the cylindrical
coordinate system fixed on a center ion of its ion atmosphere. Since the field
is axially symmetric around the $x$ axis, cylindrical coordinates are natural
coordinates to choose; see Fig. 1. Let the radial coordinate and azimuthal
angle in the plane perpendicular to the field axis be denoted $\rho$ and
$\theta$, respectively. Because of the axial symmetry of the field, the
solution of the NS equation\cite{landau} then is independent of angle $\theta$
and hence is a function of $x$ and $\rho$ only.

In the Onsager theory of electrolyte conductance\cite{onsager} the local field
in the ionic solution subjected to the external field can be calculated from
the Onsager--Fuoss (OF) equations\cite{fuoss} and the Poisson
equation\cite{jackson}, which are coupled to each other. The so-obtained local
field is then used in the NS equation (Stokes equation in fact) for the
calculation of velocity of the medium in the external field, and it provides
the spatial profiles of the countercurrent to the ionic migration in the
electrolyte solution under the influence of the external electric field, as
shown in the previous paper\cite{euIonic,archived} of this series.
Consistently with the velocity, the local nonequilibrium excess pressure is
simultaneously obtained from the NS equation. The formal Fourier transform
solutions for velocity and pressure have been explicitly calculated as
functions of position variables in a previous paper\cite{euIonic,archived}. An
application of the result for the velocity to study the equivalent conductance
of binary electrolyte solutions\cite{eucond} is in progress, especially, with
regard to the Wien effect\cite{wien,wilson,kim}---a nonlinear (non-Coulombic)
field dependence of conductance.

\section{Local Nonequilibrium Pressure}

The local nonequilibrium pressure is given in the cylindrical coordinates
$\left(  r,\rho,\theta\right)  $ by the formula\cite{euIonic,archived}
\begin{equation}
\Delta p=p-p_{0}=\frac{zeX\kappa^{2}}{4\pi^{2}}\Delta\widehat{p}\left(
r,\rho\right)  , \label{1}%
\end{equation}
where $p$ is the pressure, $p_{0}$ is the equilibrium (hydrostatic) pressure,
$\Delta\widehat{p}$ is the reduced excess nonequilibrium pressure, which is
independent of angle $\theta$ because of the cylindrical symmetry of the
system. Other symbols are: $z$ is the absolute value of the charge number of
the binary electrolyte, $e$ is the electric charge, $\kappa$ is the inverse
Debye length (Debye parameter)%
\begin{equation}
\kappa=\sqrt{\frac{4\pi e^{2}n}{Dk_{B}T}\sum_{k}z_{k}^{2}c_{k}}\qquad\left(
c_{k}=\frac{n_{k}}{n};\text{ mole fraction of }k\right)  \label{2}%
\end{equation}
with $k_{B}$ denoting the Boltzmann constant, $T$ the absolute temperature,
$n$ the density, and $D$ the dielectric constant, and finally the excess
reduced nonequilibrium pressure is given by the formula%
\begin{align}
\Delta\widehat{p}\left(  \overline{x},r\right)   &  =-\frac{\pi\overline{x}%
}{2\left(  \overline{x}^{2}+r^{2}\right)  ^{3/2}}\nonumber\\
&  \quad-\frac{\pi}{4}\int_{0}^{\sqrt{2\left(  1+\xi^{2}\right)  }}%
dy\frac{e^{-\overline{x}y}y\left(  1+\sqrt{2}\xi y+\sqrt{1+2\xi^{2}y^{2}%
}\right)  }{1+2\xi^{2}y^{2}}I_{0}(\overline{\omega}_{1}r)\nonumber\\
&  \quad+\frac{\pi}{2}\int_{0}^{1}dy\frac{e^{-\overline{x}y}\sqrt{2}\xi
y^{2}\left(  1-\sqrt{2}\xi y\right)  }{1+2\xi^{2}y^{2}}I_{0}(\overline{\omega
}_{3}r). \label{3}%
\end{align}
In this expression, the axial coordinate $\overline{x}$ and the radial
coordinate $r$ are reduced distances scaled by $\sqrt{2}\kappa^{-1}$
\begin{align}
\xi &  =\frac{zeX}{k_{B}T\kappa},\quad\overline{x}=\kappa x/\sqrt{2},\quad
r=\kappa\rho/\sqrt{2},\label{4}\\
\overline{\omega}_{1}  &  =\sqrt{1-y^{2}+\sqrt{1+2\xi^{2}y^{2}}}%
,\qquad\overline{\omega}_{3}=\sqrt{1-y^{2}}, \label{5}%
\end{align}
and $I_{0}(s)$ is the zeroth order regular Bessel
function\cite{watson,abramowitz} of second kind of argument $s$; it tends to
unity as $s\rightarrow0$, but grow exponentially as $s\rightarrow\infty$.
Therefore as the\ reduced field strength $\xi$ increases, $I_{0}%
(\overline{\omega}_{i}r)$ increases exponentially with respect to $\xi$.
Henceforth for the sake of notational brevity the overbar in the reduced
variable $\overline{x}$ will be omitted and by $x$ the reduced distance
defined in Eq. (\ref{4}) will be understood.

Therefore in some regions of $\left(  x,r\right)  $ the integrals grow
exponentially with respect to $\xi$ as will be graphically demonstrated later.
It is also easy to show that the estimates of the integrals indeed grow
exponentially with respect to $\xi$:%
\begin{equation}
\overline{\left\langle \Delta p\right\rangle }\leq\frac{e^{\sqrt{2\left(
1+\xi^{2}\right)  }R}}{\sqrt{2\pi}R^{3/2}}E\left(  \xi\right)  , \label{5est}%
\end{equation}
where%
\begin{equation}
E\left(  \xi\right)  =\int_{0}^{\sqrt{2\left(  1+\xi^{2}\right)  }}%
dy\frac{\left(  1+\sqrt{2}\xi y+\sqrt{1+2\xi^{2}y^{2}}\right)  }{\left(
1+2\xi^{2}y^{2}\right)  \left(  1-y^{2}+\sqrt{1+2\xi^{2}y^{2}}\right)  ^{1/4}%
}. \label{5E}%
\end{equation}

The integrals in Eq. (\ref{3}) are easily computed numerically, and
$\Delta\widehat{p}$ shows a negative region in the neighborhood of the
coordinate origin for all values of $\xi$, becoming negative infinite as the
origin is approached, as will be shown in Fig. 2 below.

Since ions are spherical and their force fields are spherically symmetric, for
the investigation in mind it is convenient to cast the pressure in spherical
coordinates. The cylindrical coordinates are related to the spherical
coordinates by the relations%
\begin{align}
\theta &  =\varphi,\nonumber\\
x  &  =R\cos\vartheta,\label{6}\\
r  &  =R\sin\vartheta,\nonumber
\end{align}
where $R$, $\vartheta$, and $\varphi$ are the radial, polar, and azimuthal
angle coordinates (in reduced units) in the spherical coordinate system. Owing
to $\Delta\widehat{p}$ being scalar, the spherical coordinate representation
of $\Delta\widehat{p}\left(  R,\vartheta\right)  $ is simply given by the
formula%
\begin{align}
\Delta\widehat{p}\left(  R,\vartheta\right)   &  =-\frac{\pi\cos\vartheta
}{2R^{2}}\nonumber\\
&  \quad-\frac{\pi}{4}\int_{0}^{\sqrt{2\left(  1+\xi^{2}\right)  }}%
dy\frac{e^{-yR\cos\vartheta}y\left(  1+\sqrt{2}\xi y+\sqrt{1+2\xi^{2}y^{2}%
}\right)  }{1+2\xi^{2}y^{2}}I_{0}(\overline{\omega}_{1}R\sin\vartheta
)\nonumber\\
&  \quad+\frac{\pi}{2}\int_{0}^{1}dy\frac{e^{-yR\cos\vartheta}\sqrt{2}\xi
y^{2}\left(  1-\sqrt{2}\xi y\right)  }{1+2\xi^{2}y^{2}}I_{0}(\overline{\omega
}_{3}R\sin\vartheta). \label{7}%
\end{align}
The behavior of this excess nonequilibrium\ pressure is shown in Fig. 2. This
pressure is divergent in the region in the immediate neighborhood of the
origin, but outside the region, it has also a positive region, and it means
that the ion in the neighborhood of the origin is compressed by the divergent
force toward the center by the medium under the influence of the external
electric field. We would like to examine a quantitative measure of such
compression and its relation to heat. In this connection, we remark that a
large heat emission was observed by Eckstrom and Schmeltzer\cite{eckstrom}
during their conductance experiment on electrolyte solutions.

We now calculate the excess nonequilibrium pressure on the unit area of the
surface of a sphere of radius $R$. The first term on the right of Eq.
(\ref{7}) vanishes on integration of the surface. It should be also noted that
the integrals are not uniformly convergent to a finite value in the entire
region of $R$. So, the parameter $R$ should not be taken equal to zero within
the integrals before fully evaluating them. Therefore we find%
\begin{align}
\left\langle \Delta p\right\rangle  &  =-\frac{zeX\kappa^{2}}{8\pi}\int
_{0}^{\sqrt{2\left(  1+\xi^{2}\right)  }}dy\frac{y\left(  1+\sqrt{2}\xi
y+\sqrt{1+2\xi^{2}y^{2}}\right)  }{1+2\xi^{2}y^{2}}\times\nonumber\\
&  \qquad\qquad\quad\;\int_{0}^{\pi}d\vartheta\sin\vartheta e^{-yR\cos
\vartheta}I_{0}(\overline{\omega}_{1}R\sin\vartheta)\nonumber\\
&  \qquad+\frac{zeX\kappa^{2}}{4\pi}\int_{0}^{1}dy\frac{\sqrt{2}\xi
y^{2}\left(  1-\sqrt{2}\xi y\right)  }{1+2\xi^{2}y^{2}}\times\nonumber\\
&  \qquad\qquad\quad\;\int_{0}^{\pi}d\vartheta\sin\vartheta e^{-yR\cos
\vartheta}I_{0}(\overline{\omega}_{3}R\sin\vartheta).\label{9}%
\end{align}
The two-dimensional quadratures are fairly easy to compute numerically. The
results of computation are presented in Fig. 3, in which we have plotted a
reduced $\left\langle \Delta p\right\rangle \ $vs. log$\xi$:
\begin{align}
\Psi/\xi &  =-\frac{\left\langle \Delta p\right\rangle }{8\pi k_{B}T\kappa
^{3}\xi}\nonumber\\
&  =\int_{0}^{\sqrt{2\left(  1+\xi^{2}\right)  }}dy\frac{y\left(  1+\sqrt
{2}\xi y+\sqrt{1+2\xi^{2}y^{2}}\right)  }{1+2\xi^{2}y^{2}}\times\nonumber\\
&  \qquad\qquad\quad\;\int_{0}^{\pi}d\vartheta\sin\vartheta e^{-yR\cos
\vartheta}I_{0}(\overline{\omega}_{1}R\sin\vartheta)\nonumber\\
&  \quad-\int_{0}^{1}dy\frac{2\sqrt{2}\xi y^{2}\left(  1-\sqrt{2}\xi y\right)
}{1+2\xi^{2}y^{2}}\times\nonumber\\
&  \qquad\qquad\quad\;\int_{0}^{\pi}d\vartheta\sin\vartheta e^{-yR\cos
\vartheta}I_{0}(\overline{\omega}_{3}R\sin\vartheta).\label{10}%
\end{align}
Note that here $\Psi$ is defined by
\begin{equation}
\Psi=-\frac{\left\langle \Delta p\right\rangle }{8\pi k_{B}T\kappa^{3}%
}.\label{10psi}%
\end{equation}
Since $\left\langle \Delta p\right\rangle $ is the force exerted on unit area
of the surface of radius $R$, in fact $\Psi$ is a reduced work relative to
$k_{B}T$ to compress the fluid to volume $\kappa^{-3}$ of a sphere by the
external field in which the ion is confined. As is evident from Fig. 2,
$\left\langle \Delta p\right\rangle $ is negative, and it can be deduced from
Fig. 3 that $\left\langle \Delta p\right\rangle $ tends to $\exp\left(
c\xi\right)  $ $\left(  c>0\right)  $ as the field strength increases in
confirmation of the theoretical estimate given in Eq. (\ref{5est}). Since the
bound of $\Psi$ is deduced from Eq. (\ref{5est})%
\begin{equation}
\left\vert \Psi\right\vert \leq\frac{E\left(  \xi\right)  }{\sqrt{2\pi}%
R^{3/2}}\exp\left[  R\sqrt{2\left(  1+\xi^{2}\right)  }\right]  \label{est}%
\end{equation}
and it can be deduced numerically that%
\[
\lim_{\xi\rightarrow\infty}\xi E\left(  \xi\right)  =\text{finite but not
zero}.
\]

The parameters $R=0.001$, etc. chosen for the plot in Fig. 3 are the reduced
radii of the confining space. This plot shows that $\left\langle \Delta
p\right\rangle $ gets more and more compressive as the value of $R$ decreases.
Note that $\xi=zeX/k_{B}T\kappa$ and reduced distance $R$ is defined by
$R=\kappa\widehat{R}/\sqrt{2}$, where $\widehat{R}$ is the radial coordinate
in the spherical coordinate system in actual units. We also present Table 1
for the $\Psi/\xi$ values used for Fig. 3 in the high end regime of $\xi$. We
remark that for electrolyte solutions of concentration on the order of
$10^{-3}$ mole/liter at normal temperature, the Debye radius $\kappa^{-1}$ is
on the order of $30$ \AA .

By using $\left\langle \Delta p\right\rangle $, it is possible to estimate the
molar heat emission in the electrolyte solution in an adiabatic condition
(constant entropy $S$) that accompanies the applied field:%
\begin{equation}
\left(  \Delta h\right)  _{S}=\overline{v}\left\langle \Delta p\right\rangle
_{S},\label{11}%
\end{equation}
where $\overline{v}$ is the molar volume of the solution. Because
$\left\langle \Delta p\right\rangle <0$ in general, a heat is generated by the
compression effect of the field, and it appears to explain the phenomenon of
excessive heat generation observed by Eckstrom and Schmeltzer\cite{eckstrom}
during their conductance experiment. This heat is obviously different from the
heat for the Coulomb heating effect, but a nonequilibrium thermodynamic
effect, which is probably the underlying cause for the claim of cold fusion.

\section{Discussion and Concluding Remarks}

The NS equation for velocity of the medium in binary electrolyte solutions
subjected to an external electric field can be exactly solved by using the
local potentials provided by the solutions of the coupled differential
equations of Onsager--Fuoss equations\cite{fuoss} for ionic pair distribution
functions and the Poisson equations\cite{jackson} for potentials according to
the Onsager--Wilson theory of electrolyte conductance\cite{wilson}. In the
present article, on the basis of the solution of the NS equation presented in
a previous article\cite{euIonic} we have made a deduction on the possibility
of\ local confinement of ions by means of an electric field on the basis of
the solution of the NS equation for binary electrolyte solutions. More
specifically, the nonequilibrium pressure obtained from the NS solution can be
shown to be compressive and thus emissive of heat, and this compressive force
presents a theoretical possibility of locally confining ions to a small space
by an external electric field. This compressive force is asymptotically
proportional to a product of field times an exponential function of the field
as shown in Fig. 2. This local confinement effect is an outcome of collective
motion of ions and their ion atmosphere and their mutual interactions with
each other and with the external electric field, producing a countercurrent
(velocity) to the ionic motions in the field which is nonuniformly distributed
in space. It is not clear to what extent of field strength the present theory
is applicable to electrolyte solutions before other complicating processes may
set in. Only experiment can explore the range.

It should be noted that the velocity of the countercurrent is nonuniformly
distributed in space, being singular in a region and finite in another,
especially in the periphery of ion atmosphere and thus producing a finite
electrophoretic effect. Thus in this latter region the ions are mobile and
produce conduction.\cite{eucond} Thus the fact that some ions are locally
confined does not contradict that the electrolyte solutions under the
influence of the external electric field conduct electrical currents.

\textbf{Acknowledgement}

This work was in part supported by the Natural Sciences and Engineering
Research Council of Canada through Discovery grants. The author also would
like to thank Dr. Hui Xu for assistance in drawing figures.

Figure Captions

Fig. 1. \quad The coordinate system.

Fig. 2. \quad The reduced excess nonequilibrium pressure cut at reduced radius
$R=0.001$ in spherical coordinates for the field strength $\xi=1$. At this
field strength $\Delta p\equiv\Delta\widehat{p}\left(  R,\vartheta\right)  $
is negative except for two regions of $\vartheta$ and the range of
$R\,\gtrsim1$ . $\Delta p$ gets more and more negative as $\xi$ increases.

Fig. 3. \quad Reduced force on the surface of the sphere of radius $R$ as a
function of $\xi$ for a few cases of $R$. The reduced force $\Psi$ is defined
by $\Psi=-\left\langle \Delta p\right\rangle /8\pi k_{B}T\kappa^{3}$. The
reduced force rises exponentially as $\xi$ increases.\newpage%

\begin{figure}
[ptb]
\begin{center}
\includegraphics[
natheight=5.674900in,
natwidth=6.927200in,
height=2.495in,
width=3.039in
]%
{F:/Confined 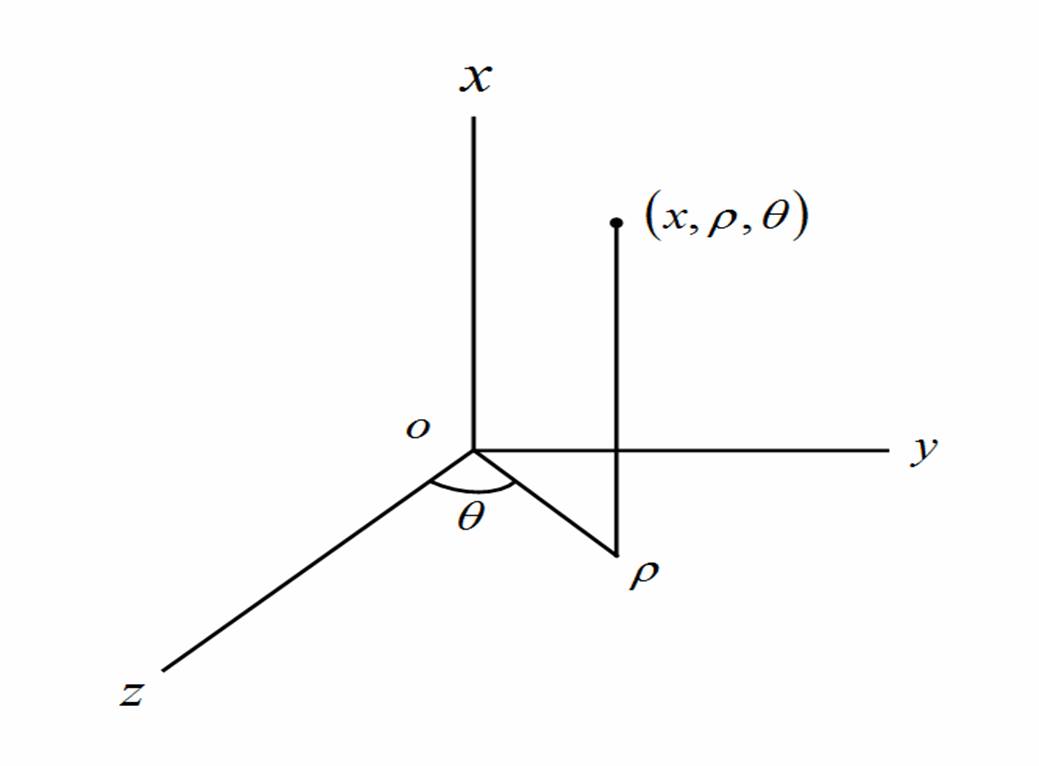}%
\caption{ }%
\label{figure1}%
\end{center}
\end{figure}
\newpage%

\begin{figure}
[ptb]
\begin{center}
\includegraphics[
natheight=6.489500in,
natwidth=7.792000in,
height=2.5356in,
width=3.039in
]%
{F:/Confined Ion/Figures/Figure2.jpg}%
\caption{ }%
\label{figure2}%
\end{center}
\end{figure}
\newpage%

\begin{figure}
[ptb]
\begin{center}
\includegraphics[
natheight=3.947900in,
natwidth=6.167000in,
height=1.9553in,
width=3.039in
]%
{F:/Confined Ion/Figures/Figure3.jpg}%
\caption{ }%
\label{figure3}%
\end{center}
\end{figure}
\newpage

\begin{center}

Table1\newline%
\begin{table}[tbp] \centering
\caption{Negative reduced pressure vs $\xi$}%
\begin{tabular}
[c]{|ll|ll|ll|}\hline\hline
$R=0.1$ &  & $R=0.01$ &  & $R=0.001$ & \\\hline
$\xi$ & \multicolumn{1}{|l|}{$\Psi/\xi$} & $\xi$ & \multicolumn{1}{|l|}{$\Psi
/\xi$} & $\xi$ & \multicolumn{1}{|l|}{$\Psi/\xi$}\\\hline\hline
$1$ & \multicolumn{1}{|l|}{$4.6732$} & $1$ & \multicolumn{1}{|l|}{$4.6707$} &
$1$ & \multicolumn{1}{|l|}{$4.6707$}\\\hline
$4$ & \multicolumn{1}{|l|}{$1.2094$} & $30$ & \multicolumn{1}{|l|}{$0.1656$} &
$400$ & \multicolumn{1}{|l|}{$0.01256$}\\\hline
$7$ & \multicolumn{1}{|l|}{$0.7099$} & $60$ & \multicolumn{1}{|l|}{$0.08416$}
& $800$ & \multicolumn{1}{|l|}{$0.00642$}\\\hline
$10$ & \multicolumn{1}{|l|}{$0.5112$} & $90$ & \multicolumn{1}{|l|}{$0.05738$}
& $1200$ & \multicolumn{1}{|l|}{$0.004448$}\\\hline
$13$ & \multicolumn{1}{|l|}{$0.4092$} & $120$ & \multicolumn{1}{|l|}{$0.04440$%
} & $1600$ & \multicolumn{1}{|l|}{$0.003533$}\\\hline
$16$ & \multicolumn{1}{|l|}{$0.3490$} & $150$ & \multicolumn{1}{|l|}{$0.03703$%
} & $2000$ & \multicolumn{1}{|l|}{$0.003062$}\\\hline
$19$ & \multicolumn{1}{|l|}{$0.3119$} & $180$ & \multicolumn{1}{|l|}{$0.03255$%
} & $2400$ & \multicolumn{1}{|l|}{$0.002838$}\\\hline
$22$ & \multicolumn{1}{|l|}{$0.2901$} & $210$ & \multicolumn{1}{|l|}{$0.02984$%
} & $2800$ & \multicolumn{1}{|l|}{$0.00279$}\\\hline
$25$ & \multicolumn{1}{|l|}{$0.2792$} & $240$ & \multicolumn{1}{|l|}{$0.02836$%
} & $3200$ & \multicolumn{1}{|l|}{$0.002899$}\\\hline
$28$ & \multicolumn{1}{|l|}{$0.2776$} & $270$ & \multicolumn{1}{|l|}{$0.02785$%
} & $3600$ & \multicolumn{1}{|l|}{$0.003176$}\\\hline
$31$ & \multicolumn{1}{|l|}{$0.2846$} & $300$ & \multicolumn{1}{|l|}{$0.02823$%
} & $4000$ & \multicolumn{1}{|l|}{$0.00366$}\\\hline
$34$ & \multicolumn{1}{|l|}{$0.3006$} & $330$ & \multicolumn{1}{|l|}{$0.02950$%
} & $4400$ & \multicolumn{1}{|l|}{$0.004421$}\\\hline
$37$ & \multicolumn{1}{|l|}{$0.3269$} & $360$ & \multicolumn{1}{|l|}{$0.03175$%
} & $4800$ & \multicolumn{1}{|l|}{$0.005569$}\\\hline
$40$ & \multicolumn{1}{|l|}{$0.3654$} & $390$ & \multicolumn{1}{|l|}{$0.03516$%
} & $5200$ & \multicolumn{1}{|l|}{$0.007278$}\\\hline
$43$ & \multicolumn{1}{|l|}{$0.4194$} & $420$ & \multicolumn{1}{|l|}{$0.03100$%
} & $5600$ & \multicolumn{1}{|l|}{$0.009811$}\\\hline
$46$ & \multicolumn{1}{|l|}{$0.4932$} & $450$ & \multicolumn{1}{|l|}{$0.04665$%
} & $6000$ & \multicolumn{1}{|l|}{$0.01358$}\\\hline
$49$ & \multicolumn{1}{|l|}{$0.5938$} & $480$ & \multicolumn{1}{|l|}{$0.05569$%
} & $6400$ & \multicolumn{1}{|l|}{$0.01920$}\\\hline
$52$ & \multicolumn{1}{|l|}{$0.7281$} & $510$ & \multicolumn{1}{|l|}{$0.06785$%
} & $6800$ & \multicolumn{1}{|l|}{$0.02763$}\\\hline\hline
\end{tabular}
\label{Table1}%
\end{table}%

\newpage

Table 1 (continued)\newline%
\begin{table}[tbp] \centering
\caption{Negative reduced pressure vs $\xi$}%
\begin{tabular}
[c]{|ll|ll|ll|}\hline\hline
$R=0.1$ &  & $R=0.01$ &  & $R=0.001$ & \\\hline
$\xi$ & \multicolumn{1}{|l|}{$\Psi/\xi$} & $\xi$ & \multicolumn{1}{|l|}{$\Psi
/\xi$} & $\xi$ & \multicolumn{1}{|l|}{$\Psi/\xi$}\\\hline\hline
$55$ & \multicolumn{1}{|l|}{$0.9085$} & $540$ & \multicolumn{1}{|l|}{$0.08419$%
} & $7200$ & \multicolumn{1}{|l|}{$0.04036$}\\
$58$ & \multicolumn{1}{|l|}{$1.1510$} & $570$ & \multicolumn{1}{|l|}{$0.1062$}
& $7600$ & \multicolumn{1}{|l|}{$0.05971$}\\\hline
$61$ & \multicolumn{1}{|l|}{$1.4776$} & $600$ & \multicolumn{1}{|l|}{$0.1357$}
& $8000$ & \multicolumn{1}{|l|}{$0.08929$}\\\hline
$64$ & \multicolumn{1}{|l|}{$1.9221$} & $630$ & \multicolumn{1}{|l|}{$0.1757$}
& $8400$ & \multicolumn{1}{|l|}{$0.1348$}\\\hline
$67$ & \multicolumn{1}{|l|}{$2.5212$} & $660$ & \multicolumn{1}{|l|}{$0.2298$}
& $8800$ & \multicolumn{1}{|l|}{$0.2050$}\\\hline
$70$ & \multicolumn{1}{|l|}{$3.3360$} & $690$ & \multicolumn{1}{|l|}{$0.3033$}
& $9200$ & \multicolumn{1}{|l|}{$0.3142$}\\\hline
$73$ & \multicolumn{1}{|l|}{$4.4480$} & $720$ & \multicolumn{1}{|l|}{$0.4036$}
& $9600$ & \multicolumn{1}{|l|}{$0.4845$}\\\hline
$76$ & \multicolumn{1}{|l|}{$5.9707$} & $750$ & \multicolumn{1}{|l|}{$0.5409$}
& $10000$ & \multicolumn{1}{|l|}{$0.7516$}\\\hline
$79$ & \multicolumn{1}{|l|}{$8.0773$} & $780$ & \multicolumn{1}{|l|}{$0.7293$}
& $10400$ & \multicolumn{1}{|l|}{$1.1720$}\\\hline
$82$ & \multicolumn{1}{|l|}{$10.966$} & $810$ & \multicolumn{1}{|l|}{$0.9889$}
& $10800$ & \multicolumn{1}{|l|}{$1.8363$}\\\hline
$85$ & \multicolumn{1}{|l|}{$14.961$} & $840$ & \multicolumn{1}{|l|}{$1.3476$}
& $11200$ & \multicolumn{1}{|l|}{$2.8901$}\\\hline
$88$ & \multicolumn{1}{|l|}{$20.501$} & $870$ & \multicolumn{1}{|l|}{$1.8446$}
& $11600$ & \multicolumn{1}{|l|}{$4.5672$}\\\hline
$91$ & \multicolumn{1}{|l|}{$28.257$} & $900$ & \multicolumn{1}{|l|}{$2.5358$}
& $12000$ & \multicolumn{1}{|l|}{$7.2451$}\\\hline
$94$ & \multicolumn{1}{|l|}{$39.02$} & $930$ & \multicolumn{1}{|l|}{$3.4987$}
& $12400$ & \multicolumn{1}{|l|}{$11.5336$}\\\hline
$97$ & \multicolumn{1}{|l|}{$54.067$} & $960$ & \multicolumn{1}{|l|}{$4.8451$}
& $12800$ & \multicolumn{1}{|l|}{$18.4213$}\\\hline
$100$ & \multicolumn{1}{|l|}{$75.155$} & $990$ & \multicolumn{1}{|l|}{$6.7313$%
} & $13200$ & \multicolumn{1}{|l|}{$29.5126$}\\\hline\hline
\end{tabular}
\label{Table1b}%
\end{table}%

\end{center}

\end{document}